# Limits on Extensions of the Minimal Standard Model from Combined LEP Lineshape Data

André Sopczak[1]

PPE Division, CERN
CH-1211 Geneva 23

### Abstract

The high statistics of the combined LEP lineshape data are used to derive constraints on hypothetical extensions of the Minimal Standard Model. The data comprises about eight million visible Z decays, recorded between 1989 and 1993. This letter gives limits for simple tests on models which predict additional Z boson decays or modified Z-couplings. As an application the two-doublet Higgs model is considered.

---

[1]E-mail: andre@cernvm.cern.ch

# Introduction

Severe limits on 'New Physics' beyond the Minimal Standard Model (MSM) [1] can be obtained from precision measurements of the Z parameters. Any hypothetical Z decay into new particles Z → X (Fig. 1a), radiative contributions from non-MSM virtual particles (Fig. 1b), or modifications to the MSM Z-couplings (Fig. 1c) are constrained by measurements of the total Z width $\Gamma_Z$, the invisible Z width $\Gamma_Z^{inv}$, the leptonic widths $\Gamma_Z^{ee}$, $\Gamma_Z^{\mu\mu}$, $\Gamma_Z^{\tau\tau}$, or the ratio of the hadronic to leptonic Z decay width $R$. Thus, constraints on physics beyond the MSM can be expressed as limits on deviations from the MSM Z decay width predictions. In particular, such limits can be used to constrain the existence of Higgs bosons in models with more than one Higgs doublet; charginos, neutralinos and light gluinos in Supersymmetric Models with or without R-parity conservation; additional heavy charged or neutral leptons; or anomalous gauge boson couplings.

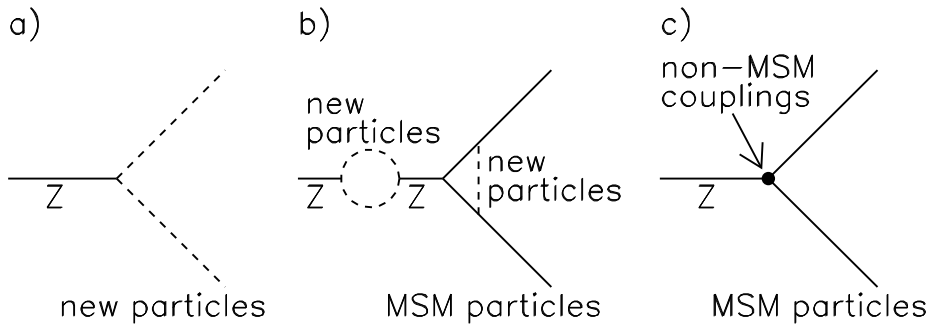

Figure 1: Illustration of possible effects in extensions of the MSM which can be constrained by a comparison of measured Z parameters with MSM predictions.

The present analysis includes 1992 data [2] and preliminary 1993 data collected by the four LEP experiments [3], corresponding to a total of about eight million visible Z decays. The Z parameters are obtained by fitting the lineshape of the Z decay into charged leptons and hadrons. All measurements are in agreement with the MSM predictions. Details of the experimental analysis and similar interpretations as presented here can be found in the corresponding publications of the four LEP experiments [2].

# Measurement and Theory

Table 1 summarizes the measured values of $\Gamma_Z$, $\Gamma_Z^{inv}$, $\Gamma_Z^{ee}$, $\Gamma_Z^{\mu\mu}$, $\Gamma_Z^{\tau\tau}$, and R, as well as their MSM upper and lower bounds for one-sided 95% CL's. One-sided CL's are used because a new decay would always increase the Z width; they are derived assuming Gaussian errors by extending the 1 $\sigma$ error to 1.64 $\sigma$ [4]. The measured values are averages from the four LEP experiments taking into account common systematic errors [3]. Theoretical upper and lower bounds are obtained with an analytical program (ZFITTER version 4.6 [5]) by varying the strong coupling constant $\alpha_s$, the top quark mass $m_t$, and the MSM Higgs mass $m_h$, independently within their one-sided 95% CL limits. The uncertainty in these values constitutes the dominant error on the MSM predictions.

For $\alpha_s$ the world average $\alpha_s(m_Z) = 0.125 \pm 0.005$ [3] is used. Note that this average is based on data from $\nu$N experiments, p$\bar{\text{p}}$ colliders, the SLD measurement of the left-right asymmetry and the LEP experiments. For $m_t$ the limit implied by the recently reported CDF evidence for the top quark is used, i.e. $m_t = (174 \pm 10^{+13}_{-12})$ GeV [6]. For $m_h$ a combined lower mass limit [7], resulting from the data of the four LEP experiments [8], and a theoretical upper mass bound following from consistency arguments in the MSM [9] is used. Thus, the ranges used for $\alpha_s$, $m_t$ and $m_h$ are:

$$0.117 < \alpha_s(m_Z) < 0.133, \quad (148 < m_t < 201) \text{ GeV}, \quad (63.5 < m_h < 1000) \text{ GeV}.$$

The central values of the MSM predictions are the arithmetic means of the upper and lower bounds.

| Parameter | Measurements | | | Theory (MSM) | | |
|---|---|---|---|---|---|---|
| | Mean Value | Lower Bound | Upper Bound | Lower Bound | Upper Bound | Mean Value |
| $\Gamma_Z$ | $2497.4 \pm 3.8$ | 2491.2 | 2503.6 | 2480.6 | 2512.3 | 2496.5 |
| $\Gamma_Z^{\text{inv}}$ | $499.8 \pm 3.5$ | 494.1 | 505.5 | 499.7 | 503.4 | 501.6 |
| $\Gamma_Z^{ee}$ | $83.85 \pm 0.21$ | 83.51 | 84.19 | 83.56 | 84.33 | 83.95 |
| $\Gamma_Z^{\mu\mu}$ | $83.95 \pm 0.30$ | 83.46 | 84.44 | 83.56 | 84.33 | 83.95 |
| $\Gamma_Z^{\tau\tau}$ | $84.26 \pm 0.34$ | 83.70 | 84.82 | 83.37 | 84.13 | 83.75 |
| $R$ | $20.795 \pm 0.040$ | 20.729 | 20.861 | 20.692 | 20.842 | 20.767 |

Table 1: Measured Z parameters, MSM predictions, and their lower and upper limits at one-sided 95% CL's. All decay widths are given in MeV.

In order to obtain a conservative limit on non-MSM effects from $\Gamma_Z$, one considers the intervals:

$$(\Gamma_Z)^{\text{exp}}_{\text{min}} - (\Gamma_Z)^{\text{th}}_{\text{max}} \quad \text{and} \quad (\Gamma_Z)^{\text{exp}}_{\text{max}} - (\Gamma_Z)^{\text{th}}_{\text{min}},$$

where both the experimental and theoretical limits are taken at the one-sided 95% CL. Similar intervals are defined for the other parameters listed in Table 1. If the value of the predicted mean value minus the measured mean value is negative (positive), it is added to the lower (upper) limit. This conservative approach avoids setting tighter constraints than allowed by the agreement between theory and measurement. Otherwise, e.g., a measurement of the central value of the total Z decay width significantly below the MSM expectation would naively lead to a too strong bound on physics processes beyond the MSM. Table 2 summarizes the intervals and differences obtained.

| Parameter | Interval | Difference | Sum |
|---|---|---|---|
| $(\delta\Gamma_Z)_{\min}$ | $-21.1$ | $-0.9$ | $-22.0$ |
| $(\delta\Gamma_Z)_{\max}$ | $23.0$ | $0$ | $23.0$ |
| $(\delta\Gamma_Z^{\rm inv})_{\min}$ | $-9.3$ | $0$ | $-9.3$ |
| $(\delta\Gamma_Z^{\rm inv})_{\max}$ | $5.8$ | $1.8$ | $7.6$ |
| $(\delta\Gamma_Z^{\rm ee})_{\min}$ | $-0.82$ | $0$ | $-0.82$ |
| $(\delta\Gamma_Z^{\rm ee})_{\max}$ | $0.63$ | $0.10$ | $0.73$ |
| $(\delta\Gamma_Z^{\mu\mu})_{\min}$ | $-0.87$ | $0$ | $-0.87$ |
| $(\delta\Gamma_Z^{\mu\mu})_{\max}$ | $0.88$ | $0$ | $0.88$ |
| $(\delta\Gamma_Z^{\tau\tau})_{\min}$ | $-0.43$ | $-0.51$ | $-0.94$ |
| $(\delta\Gamma_Z^{\tau\tau})_{\max}$ | $1.45$ | $0$ | $1.45$ |
| $(\delta R)_{\min}$ | $-0.113$ | $-0.028$ | $-0.141$ |
| $(\delta R)_{\max}$ | $0.169$ | $0$ | $0.169$ |

Table 2: Allowed changes of $\Gamma_Z$, $\Gamma_Z^{\rm inv}$, $\Gamma_Z^{\rm ee}$, $\Gamma_Z^{\mu\mu}$, $\Gamma_Z^{\tau\tau}$, and R due to non-MSM contributions, using twice one-sided 95% CL limits. Max indicates the maximum experimental value minus the minimum theoretical value, and min indicates the minimum experimental value minus the maximum theoretical value. The interval and difference are defined in the text. All decay widths are given in MeV.

Considering the new decay channel Z $\to$ X, let the decay ratios of X be defined as $x_j \equiv \Gamma(X \to j)/\Gamma(X \to {\rm anything})$, where $j = h, l, i$ for hadrons, leptons and invisible particles, respectively. In this definition, $x_h + x_l + x_i = 1$. Let the hadronic and leptonic branching ratios of the Z be $b_h$ and $b_l$, respectively. In the definition of $R$, the hadronic Z decays are summed over all five quark types produced at LEP, while the leptonic Z decay width is given for a massless charged lepton pair assuming lepton universality. Let $\Gamma_Z^X \equiv \Gamma(Z \to X)$, then

1) The limit on $\Gamma_Z^X$ from $\Gamma_Z$ is given by:

$$\Gamma_Z^X \leq (\delta\Gamma_Z)_{\max} = 23.0 \text{ MeV}. \qquad (1)$$

2) The limit on $\Gamma_Z^X$ from $\Gamma_Z^{\rm inv}$ is given by:

$$x_i \Gamma_Z^X \leq (\delta\Gamma_Z^{\rm inv})_{\max} = 7.6 \text{ MeV}. \qquad (2)$$

3) A contribution from Z $\to$ X decays would change the ratio $R = b_h/b_l$ by:

$$\delta R = \frac{\Gamma_Z b_h + \Gamma_Z^X x_h}{\Gamma_Z b_l + \frac{1}{3}\Gamma_Z^X x_l} - \frac{b_h}{b_l} \approx R\frac{\Gamma_Z^X}{\Gamma_Z}(\frac{x_h}{b_h} - \frac{x_l}{3b_l}), \qquad (3)$$

an approximation which is valid when $\Gamma_Z^X \ll \Gamma_Z$. For $x_h = 1$ and $x_l = 0$, $(\delta R)_{max}$ leads to $\Gamma_Z^X \le 14$ MeV. For $x_h = 0$ and $x_l = 1$, $(\delta R)_{min}$ results in $\Gamma_Z^X \le 1.7$ MeV, however, this limit is weaker than those from $\Gamma_Z^{ee}$, $\Gamma_Z^{\mu\mu}$, $\Gamma_Z^{\tau\tau}$. Recently, a lower limit on the gluino mass of 3.8 GeV derived from $R$-measurements has been reported at 90% CL [10].

Radiative contributions from non-MSM virtual particles or modifications to the MSM Z-couplings are constrained by the upper and lower limits given in Table 2.

## Discussion and Example

The most stringent limits on deviations from the non-MSM effects on the Z decay widths are summarized in Table 3. Both upper and lower limits are given at one-sided 95% CL. As a consequence, modified MSM Z-couplings or amplitudes of non-MSM radiative corrections are constrained to the interval at 90% CL. The limits on new decay modes obtained from $\Gamma_Z$ are independent of the decay branching fractions, while the limits from $\Gamma_Z^{inv}$ constrain only invisible Z decay modes. The limits from $\Gamma_Z^{ee}$, $\Gamma_Z^{\mu\mu}$, $\Gamma_Z^{\tau\tau}$, and $R$ constrain the corresponding leptonic and hadronic Z-couplings, respectively. The limits for unspecified and invisible decay modes are of most general use. The limits on $\Gamma_Z^{ee}$ are tighter, since the Zee-coupling contributes both to Z production and decay. One should note that the charged leptonic and hadronic limits are not able to constrain Z decays if the resulting new particles subsequently decay; dedicated searches are necessary for such specific final states. This is due to the precise selection criteria applied for leptonic and hadronic Z decay event topologies. If a model predicts the invisible, charged leptonic and hadronic branching fractions of Z decays, a $\chi^2$-method allows setting tighter constraints.

| Origin | Decay Mode | $\Delta\Gamma(Z)$ (MeV) | | $\Delta\mathrm{Br}(Z)$ (in %) |
|---|---|---|---|---|
| $\Gamma_Z$ | Z→anything | −22.0 | 23.0 | 0.92 |
| $\Gamma_Z^{inv}$ | Z→invisible | −9.3 | 7.6 | 0.30 |
| $\Gamma_Z^{ee}$ | Z→$e^+e^-$ | −0.82 | 0.73 | 0.029 |
| $\Gamma_Z^{\mu\mu}$ | Z→$\mu^+\mu^-$ | −0.87 | 0.88 | 0.035 |
| $\Gamma_Z^{\tau\tau}$ | Z→$\tau^+\tau^-$ | −0.94 | 1.45 | 0.058 |
| $R$ | Z→hadrons | −12 | 14 | 0.56 |

Table 3: One-sided 95% CL lower and upper limits on $\Delta\Gamma(Z)$ for Z decaying into any, invisible, charged leptonic, and hadronic channel. The corresponding branching ratio upper limits on $\Delta\mathrm{Br}(Z)$ are also given.

The present study updates the analyses given in [11] which were based on 1990 and 1991 LEP data. Only slightly tighter limits are obtained by including the 1992 data as they were entirely taken on the Z pole. Including the 1993 data which contain both off and on peak results, limits are significantly improved: the experimental errors are reduced by about a factor two compared to those used in [11]. In this regard, little improvement is expected from the 1994 data as they are taken again on the Z-pole only. For unspecified Z decays, the 1993 improvement of $\Delta\Gamma(Z)$ is mainly due to the increased predicted MSM lower bound on $\Gamma_Z$ following from the new top mass constraints of the CDF experiment.

As an example, a limit on $\cos^2(\beta - \alpha)$ in the general two-doublet Higgs model is derived. The definitions used are: $\tan\beta$ the ratio of the vacuum expectation values of the Higgs doublets and $\alpha$ the mixing angle between the neutral scalar Higgs fields. The Z decay width into neutral Higgs pairs in the general two-doublet Higgs model is given by [12]:

$$\Gamma(Z \to hA) = \Gamma(Z \to \nu\bar{\nu})\frac{1}{2}\cos^2(\beta - \alpha)\lambda^{3/2}(\frac{m_h^2}{m_Z^2}, \frac{m_A^2}{m_Z^2}), \tag{4}$$
$$\lambda(a,b) = (1 - a - b)^2 - 4ab,$$

with $\Gamma(Z \to \nu\bar{\nu})$ derived from a combined Z lineshape fit: $\Gamma(Z \to \nu\bar{\nu}) = 166.6 \pm 1.2$ MeV [3]. Without any assumption on the Higgs decay modes, the constraint $\Delta\Gamma(Z) \leq 23.0$ MeV sets a limit on $\cos^2(\beta - \alpha)$ as a function of $m_h$ and $m_A$:

$$\cos^2_{\max}(\beta - \alpha) = \frac{2\Gamma_Z^X}{\Gamma(Z \to \nu\bar{\nu})}\lambda^{-3/2}(\frac{m_h^2}{m_Z^2}, \frac{m_A^2}{m_Z^2}). \tag{5}$$

Figure 2 shows the excluded $\cos^2(\beta - \alpha)$ range at 95% CL as a function of $m_h$ for $m_A = 20$ GeV. In conjunction with a constraint on $\sin^2(\beta - \alpha)$, derived from the search for the MSM Higgs boson, this limit leads to an exclusion of a large $(m_h, m_A)$ parameter range [7]. Further constraints can result from an analysis of the one-loop vertex corrections to the Zbb-coupling involving additional neutral and charged Higgs bosons; such corrections could decrease $\Gamma(Z \to b\bar{b})$ and thus the hadronic decay width, depending on the unknown parameters of the two-doublet Higgs model [13]. In this case the limit $\Delta\Gamma(Z \to \text{hadrons}) \geq -12$ MeV applies.

## Acknowledgments

I would like to thank Sally Dawson, Wolfgang Hollik, Joachim Mnich, Stefan Pokorski, and Janusz Rosiek for fruitful discussions and I express my gratitude to the Institute of Theoretical Physics at the Warsaw University for their hospitality. For his advice on finalizing the manuscript I thank Remy Van de Walle.

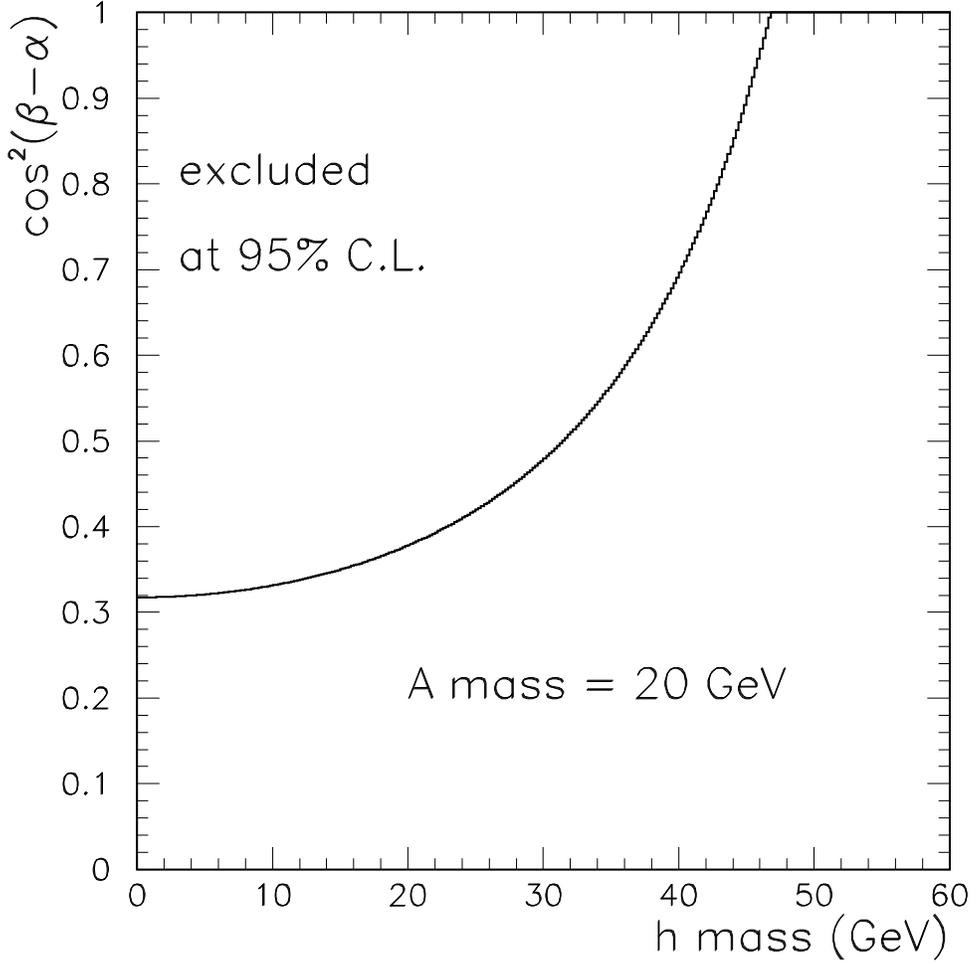

Figure 2: Limit on $\cos^2(\beta - \alpha)$ of the general two-doublet Higgs model as a function of $m_h$ for $m_A = 20$ GeV. The limit is based on the constraint $\Delta\Gamma(Z \to \text{anything}) \leq 23.0$ MeV, set by the precision lineshape measurements. No assumptions on the decay branching ratios of the Higgs bosons are made.